\documentclass[pra,twocolumn,showpacs,preprintnumbers,amsmath,amssymb]{revtex4}
\usepackage{mathrsfs}
\usepackage{bbm}
\usepackage{amsfonts}
\usepackage{tipa}

\usepackage{epsfig,graphicx}
\usepackage{amstext}
\usepackage{amsmath}
\usepackage{graphicx}
\usepackage{times}

\begin{document}
%%%%%%%%%%%%%%%%%%%%%%%%%%%%%%%%%%%%%%%%%%%%%%%%%%%%%%%%%%%%%%%%%%%%%%

%TCIDATA{OutputFilter=Latex.dll}
%TCIDATA{Version=5.00.0.2552}
%TCIDATA{<META NAME="SaveForMode" CONTENT="1">}
%TCIDATA{LastRevised=Wednesday, June 22, 2005 16:21:09}
%TCIDATA{<META NAME="GraphicsSave" CONTENT="32">}

\title{Competitions between quantum correlations in the quantum-memory-assisted entropic uncertainty relation}

\author{Ming-Liang Hu$^{1,}$}
\email{mingliang0301@163.com}
\author{Heng Fan$^{2,}$}
\email{hfan@iphy.ac.cn}
\address{$^{1}$School of Science, Xi'an University of Posts and
               Telecommunications, Xi'an 710121, China \\
         $^{2}$Beijing National Laboratory for Condensed Matter Physics,
               Institute of Physics, Chinese Academy of Sciences, Beijing
               100190, China}

\begin{abstract}
With the aid of a quantum memory, the uncertainty about the
measurement outcomes of two incompatible observables of a quantum
system can be reduced. We investigate this measurement uncertainty
bound by considering an additional quantum system connected with
both the quantum memory and the measured quantum system. We find
that the reduction of the uncertainty bound induced by a quantum
memory, on the other hand, implies its increasing for a third
participant. We also show that the properties of the uncertainty
bound can be viewed from perspectives of both quantum and classical
correlations, in particular, the behavior of the uncertainty bound
is a result of competitions of various correlations between
different parties.
\end{abstract}

\pacs{03.67.Mn, 03.65.Ta, 03.65.Yz
%\\Key Words:
}

\maketitle

\section{Introduction}
 The Heisenberg uncertainty principle \cite{Heisenberg} is one
of the most remarkable features of quantum theory which differs
quantum world essentially from the classical world. It sets limits
on the precise prediction of the outcomes of two incompatible
quantum measurements $Q$ and $R$ on a particle, and is expressed in
various forms \cite{Robertson,Deutsch,Prevedel}. However, Berta {\it
et al.} \cite{Berta} showed recently that the uncertainty bound
imposed by the Heisenberg principle could actually be violated with
the aid of a quantum memory $B$ that is entangled with the particle
$A$ to be measured. This quantum-memory-assisted entropic
uncertainty relation reads \cite{Berta}
\begin{equation}\label{eq1}
 S(Q|B)+S(R|B)\geqslant \log_2 \frac{1}{c}+S(A|B),
\end{equation}
the equivalent form of which was previously conjectured by Renes and
Boileau \cite{Renes}. Here, $S(A|B)$ is the conditional von Neumann
entropy of the density operator $\rho_{AB}$, $S(A|B)=S(\rho
_{AB})-S(\rho _B)$. On the left-hand side (LHS) of the inequality,
$S(X|B)$ is that of the postmeasurement state
$\rho_{XB}=\sum_{k}(\Pi_k^{X}\otimes
\mathbb{I})\rho_{AB}(\Pi_k^{X}\otimes \mathbb{I})$ which represents
uncertainty of the measurement outcomes of $X=\{Q,R\}$ conditioned
on the prior information stored in $B$, where
$\Pi_k^{X}=|\Psi_k^X\rangle\langle\Psi_k^X|$ with $|\Psi_k^X\rangle$
being the eigenstates of $X$, and $c=\max_{k,l}|\langle
\Psi_k^Q|\Psi_l^R\rangle|^2$ with $1/c$ quantifies the
complementarity of $Q$ and $R$.

This generalized entropic uncertainty relation has been confirmed in
all-optical experiments \cite{Prevedel,Licf}. Meanwhile, the related
relations expressed by other entropic quantities, such as the
R\'{e}nyi entropy which is important in physical models
\cite{Cuijian}, are also exploited \cite{Tomamichel,Colesprl}. Since
it has a fundamental role, this quantum-memory-assisted entropic
uncertainty relation can be studied from various viewpoints
\cite{Colesarxiv,Xuzy,Xuzhu}, and can be applied to other quantum
information processes \cite{Renner,Hupra,Pati}.

The uncertainty relation of Eq. \eqref{eq1} differs from its
original one \cite{Deutsch} by an additional term $S(A|B)$. It is
clear that the bound of the entropic uncertainty, the right-hand
side of inequality \eqref{eq1} named as the uncertainty bound (UB)
hereafter, is reduced whenever $S(A|B)<0$. It is remarkable that the
quantity of conditional entropy $S(A|B)$ has many important
implications in quantum information processing. Its negativity means
inseparability \cite{Cerf} and gives the lower bound of the one-way
distillable entanglement for $\rho_{AB}$ \cite{Devetak}. It
quantifies partial quantum information \cite{opeh} and can be
related with quantum correlation measures
\cite{Xuzy,Ollivier,Luopra,Dakic,Modi,Luomin}.

In this paper, we go one step further from bipartite state $\rho
_{AB}$ to consider its purification $|\Psi \rangle _{ABC}$ or a
tripartite state $\rho _{ABC}$, i.e., a third party $C$ is entangled
with both the particle $A$ and the memory $B$. Some fundamental and
interesting phenomena are found: For example, there exists
correlative capacities which indicate the uncertainty reduction of
UB because of $B$ implies its increasing for other parties; the
changing of UB is induced by competitions of various quantum
correlations between different pairs. These results have important
conceptual implications and shed new light on the foundations of
quantum mechanics.

\section{Correlation capacities}
We begin with a simple yet
meaningful observation. For any three-partite system $ABC$ with
density matrix $\rho_{ABC}$, we have
\begin{equation}\label{eq2}
 S(A|B)+S(A|C)\geqslant 0,
\end{equation}
which can be proved directly by the strong subadditivity inequality:
$S(\rho_B)+S(\rho_C)\leqslant S(\rho_{AB})+S(\rho_{AC})$
\cite{Nielsen}. Eq. \eqref{eq2} indicates that whenever $S(A|B)<0$,
we always have $S(A|C)>0$. Therefore, the reduction of the UB on $A$
with quantum information stored in $B$ excludes its reduction by
quantum information stored in $C$. Since the reduction of the UB
originates from the quantum correlations established between the
measured particle and the quantum memory \cite{Berta}, this
observation may be interpreted as a fact that particle $A$ reaches
its potential correlative capacities with the quantum memory $B$ in
the sense that any other quantum memory except $B$ always gives
increasing UB of measurement uncertainty on $A$.

\begin{figure}
\centering
\resizebox{0.4\textwidth}{!}{%
\includegraphics{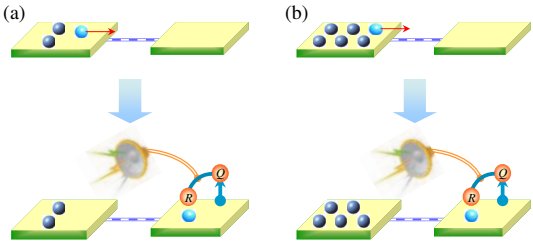}}
% If not, use\vspace{5cm}
% Give the correct figure height in cm
\caption{(Color online) Schematic representation of the
``uncertainty game" with three (a) and six (b) players.}
\label{fig:1}
% Give a unique label
\end{figure}

Inequality \eqref{eq2} also has important physical implications. To
be convinced, let us consider a variant of the imaginary
``uncertainty game" presented in \cite{Berta}: Three players, Alice
($A$), Bob ($B$), and Charlie ($C$), share a quantum state
$\rho_{ABC}$, the form of which is known only to Bob and Charlie.
They begin the game by preagreeing on two measurements $Q$ and $R$.
Alice then measures either $Q$ or $R$ randomly on the particle $A$,
and informs Bob and Charlie of her measurement choice but not the
outcome. What we want to determine here is whether Bob and Charlie
(communication between them is forbidden) can predict the outcomes
of Alice both with improved precision (see Fig. \ref{fig:1} for an
illustration). As for fixed $Q$ and $R$, the UB of measurement is
determined only by the conditional entropy $S(A|X)$ for $\rho_{AX}$
of the observed particle $A$ and the quantum memories $X=\{B,C\}$,
and Eq. \eqref{eq2} excludes the possibility for $S(A|B)$ and
$S(A|C)$ taking the negative values simultaneously, the prediction
precision of Bob and Charlie cannot be improved simultaneously in
this game; i.e., the improvement of Bob's prediction precision
implies the degradation for that of Charlie's, and vice versa.
Particularly, for pure $\rho_{AB}$, Eq. \eqref{eq2} simplifies to
$S(A|B)+S(A|C)=0$, which tells us that the more precisely the
measurement outcome is predicted by one participant, the less
precisely that will be predicted by the other one. In some sense,
one may say that this implies another kind of uncertainty relation,
because it sets limits on Bob and Charlie's (under the constraint of
no communication between them) ability to predict correctly the
measurement outcomes of Alice simultaneously; that is to say, the
certainty of prediction for one participant implies the uncertainty
of prediction for another participant

The arguments above can also be easily generalized to the $N$-player
case, i.e., we have
\begin{equation}\label{eq3}
 \sum_{i=1}^{N-1}S(A|X_i)\geqslant 0,
\end{equation}
where the particles $X_{i}=\{B,C,D,\cdots\}$ belong to Bob, Charlie
and Daniel, {\it et al.}, respectively. This inequality can be
proved directly by combination of the strong subadditivity of the
von Neumann entropy (Theorem 11.14 of \cite{Nielsen}) and the
subadditivity of the conditional entropy (Theorem 11.16 of
\cite{Nielsen}), and it implies that even for the multiplayer case,
the precision of predictions about Alice's measurement outcomes
cannot be improved simultaneously for all of the participants.

Also we would like to remark here that the generalized ``uncertainty
game" illustrated in Fig. \ref{fig:1} can be immediately tested by
similar all-optical setups as those in Refs. \cite{Prevedel,Licf}.

\section{Competition of quantum discords}
Next let us consider some
quantum correlations and begin with quantum discord \cite{Ollivier}.
The measure of the classical correlation takes the form
\begin{equation}\label{eq4}
 J(B|A)=S(\rho_B)-\min_{\{E_k^A\}}S(B|\{E_k^A\}),
\end{equation}
where $S(B|\{E_k^A\})=\sum_{k}p_{k}S(\rho_{B|E_k^A})$, with
$\rho_{B|E_k^A}={\rm Tr}_{A}(E_k^A\rho_{AB})/p_k$ being the
nonselective postmeasurement state of $B$ after the positive
operator valued measure (POVM) on $A$, and
$p_k=\text{Tr}(E_k^A\rho_{AB})$ is the probability for obtaining the
outcome $k$. $J(B|A)$ is usually interpreted as the maximum
information gained about $B$ with the measurement outcome of $A$.
The quantum discord is then defined by the discrepancy between
quantum mutual information $I(A:B)=S(\rho_A)+S(\rho_B)-S(\rho_{AB})$
and $J(B|A)$ as
\begin{equation}\label{eq5}
 D(B|A)=I(A:B)-J(B|A).
\end{equation}
The quantum discord can therefore be interpreted as the minimal loss
of correlations due to the POVM $\{E_k^A\}$. It survives for states
with quantumness of correlation and vanishes for states with only
classical correlation. It attracts much attention recently because
of its fundamental role in quantum information processing
\cite{Datta,Madhok,Dakicnp,Gunp}. Here, we demonstrate a new
perspective of quantum discord in the uncertainty principle of
quantum mechanics.

Assume $|\Psi \rangle _{ABC}$ being the purification of the
bipartite state $\rho _{AB}$, we first have the following
proposition:

\emph{Proposition 1.} When the UB on $A$ is reduced with the aid of
a quantum memory $B$, then both $D(B|A)$, $J(B|A)$ and the
entanglement of formation (EoF), $E_f(\rho_{AB})$, are larger than
those between $A$ and its purifying system $C$.

\emph{Proof.} By using the Koashi-Winter equality for
$|\Psi\rangle_{ABC}$ \cite{Koashi}, we obtain
\begin{eqnarray}\label{eq6}
 &&E_f(\rho_{BC})+J(B|A)=S(\rho_B),\nonumber\\
 &&E_f(\rho_{CB})+J(C|A)=S(\rho_C),
\end{eqnarray}
where $E_f(\rho_{BC})$ is the EoF for $\rho_{BC}$, defined as
$E_f(\rho_{BC})=\min_{\{p_i,|\psi_i\rangle_{BC}\}}\sum_i p_i S({\rm
Tr}_C |\psi_i\rangle_{BC}\langle\psi_i| )$ \cite{Bennett}, and the
minimum is taken over all pure state decompositions
$\rho_{BC}=\sum_i p_i |\psi_i\rangle_{BC}\langle\psi_i|$. Since
$E_f(\rho_{BC})=E_f(\rho_{CB})$, Eq. \eqref{eq6} yields
$J(B|A)-J(C|A)=S(\rho_B)-S(\rho_C)=-S(A|B)>0$, and therefore
$J(B|A)>J(C|A)$. Furthermore, by combining Eqs. \eqref{eq5} and
\eqref{eq6}, we obtain an equivalent form of the Koashi-Winter
equalities
\begin{eqnarray}\label{eq7}
 &&D(B|A)+S(B|A)=E_f(\rho_{BC}),\nonumber\\
 &&D(C|A)+S(C|A)=E_f(\rho_{CB}),
\end{eqnarray}
which gives $D(B|A)-D(C|A)=S(C|A)-S(B|A)=-S(A|B)>0$, and hence
$D(B|A)>D(C|A)$. Finally, to prove $E_f(\rho_{AB}>E_f(\rho_{AC})$,
we note that the conditional entropy $S(A|B)<0$ is equivalent to
$S(\rho_B)>S(\rho_C)$ for $|\Psi\rangle_{ABC}$. Therefore by using
the chain rule \cite{Giorgi}, we derive
\begin{eqnarray}\label{eq8}
  S(\rho_B)+E_f(\rho_{CA})\leqslant S(\rho_C)+E_f(\rho_{AB}),
\end{eqnarray}
which implies $E_f(\rho_{AB})-E_f(\rho_{AC})\geqslant
S(\rho_B)-S(\rho_C)=-S(A|B)>0$, and thus completes the
proof.\hfill{$\blacksquare $}

In fact, from Eqs. \eqref{eq4}, \eqref{eq5}, and
$S(\rho_{B|A})=S(\rho_{C|A})=E_f(\rho_{BC})$, with
$S(\rho_{X|A}):=\min_{\{E_k^A\}}S(X|\{E_k^A\})$ ($X=B$ or $C$)
\cite{Koashi}, we can obtain
\begin{eqnarray}\label{eq9}
 D(B|A)+J(C|A)=D(C|A)+J(B|A)=S(\rho_A).
\end{eqnarray}
Thus, $D(B|A)>D(C|A)$ and $J(B|A)>J(C|A)$ are in fact equivalent,
i.e., the fulfillment of one inequality implies the holding of
another one. Moreover, we point out here that even for mixed
$\rho_{ABC}$, we still have $J(B|A)>J(C|A)$. This is because for any
$\rho_{ABC}$ with the purification $|\Psi\rangle_{ABCD}$, we always
have $J(B|A)>J(CD|A)\geqslant J(C|A)$, where the first inequality
originates from Proposition 1 (by taking $CD$ as a combined system),
and the second one is due to the fact that the classical correlation
is nonincreasing under local quantum operations \cite{Ollivier}.

Eq. \eqref{eq8} also implies that $J(C|B)\geqslant J(B|C)$, which
can be convinced by the Koashi-Winter equalities
$J(C|B)=S(\rho_C)-E_f(\rho_{CA})$ and
$J(B|C)=S(\rho_B)-E_f(\rho_{AB})$. By combining this with
$D(A|B)+J(C|B)=S(\rho_B)$ [an equivalent form of Eq. \eqref{eq9}]
and $D(A|B)+S(A|B)=E_f(\rho_{AC})$, we further obtain
\begin{eqnarray}\label{eq10}
 E_f(\rho_{AC})<D(A|B)\leqslant E_f(\rho_{AB}).
\end{eqnarray}
This equation indicates that when the UB on $A$ is reduced, the
quantum discord $D(A|B)$ is upper bounded by EoF between $A$ and the
quantum memory $B$ and lower bounded by EoF between $A$ and the
purifying system $C$. Furthermore, for pure $|\Psi\rangle_{ABC}$ Eq.
\eqref{eq2} turns into $S(A|B)+S(A|C)=0$, therefore by combining
this with the Koashi-Winter equalities of the equivalent form of Eq.
\eqref{eq7}, we have $D(A|B)+D(A|C)=E_f(\rho_{AB})+E_f(\rho_{AC})$,
and hence Eq. \eqref{eq10} also means $E_f(\rho_{AC})\leqslant
D(A|C)< E_f(\rho_{AB})$.

We now discuss the physical mechanism responsible for changing UB.
From the proof of Proposition 1 we know that
\begin{eqnarray}\label{eq11}
 S(A|B)=D(C|A)-D(B|A),
\end{eqnarray}
and therefore $S(A|B)$ is determined by the competition between the
quantum discords $D(C|A)$ and $D(B|A)$. This relation has also been
noted by Fanchini {\it et al.} \cite{Felipe}. It explains why the UB
is not a monotonic function of the quantum discord between $A$ and
the quantum memory $B$, as while $D(B|A)$ increases, $D(C|A)$ may
also increases but with a faster rate, and as a result, this induces
the increase of the uncertainty with increasing $D(B|A)$. To be
explicit, we consider the mixed state $\rho_{AB}$ of the following
form
\begin{eqnarray}\label{eq12}
 \rho_{AB}=\sin^2\theta |\Phi\rangle\langle\Phi|+\cos^2\theta|11\rangle\langle11|,
\end{eqnarray}
where $|\Phi\rangle=\cos\phi|01\rangle+\sin\phi|10\rangle$ in the
standard basis $\{|0\rangle,|1\rangle\}$. The purification
$|\Psi\rangle_{ABC}$ for this state can be written as
\begin{eqnarray}\label{eq13}
 |\Psi\rangle_{ABC}&=&\sin\theta\cos\phi|011\rangle+\sin\theta\sin\phi|101\rangle\nonumber\\
                    &&+\cos\theta|110\rangle,
\end{eqnarray}
which is just the generalized {\it W} state \cite{Dur}.

\begin{figure}
\centering
\resizebox{0.38\textwidth}{!}{%
\includegraphics{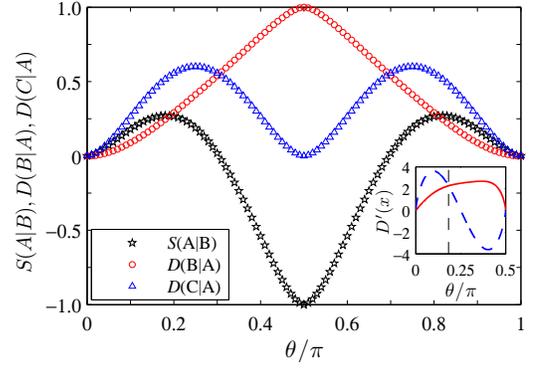}}

% If not, use\vspace{5cm}
% Give the correct figure height in cm
\caption{(Color online) Conditional von Neumann entropy
         $S(A|B)$ and quantum discords $D(B|A)$ and $D(C|A)$ versus $\theta/\pi$
         for $|\Psi\rangle_{ABC}$ of Eq. \eqref{eq13} with $\phi=\pi/4$. The insets are
         derivatives of $D(x)$ with respect to $\theta/\pi$, with $x=B|A $
         (solid red), $C|A$ (dashed blue), and the vertical dashed line represents constant
         0.182.} \label{fig:2}
% Give a unique label
\end{figure}

For the purification $|\Psi\rangle_{ABC}$ of Eq. \eqref{eq13}, both
the reduced states $\rho_{AB}$ and $\rho_{AC}$ have the X structure,
and therefore the discords $D(B|A)$ and $D(C|A)$ can be determined
analytically \cite{Luoetal}. In Fig. \ref{fig:2} we plot dependence
of $S(A|B)$, $D(B|A)$, and $D(C|A)$ on $\theta/\pi$ with
$\phi=\pi/4$, i.e., $|\Phi\rangle=(|01\rangle+|10\rangle)/\sqrt{2}$.
One can see that $S(A|B)$, and thus the UB, increases with
increasing values of both $D(B|A)$ and $D(C|A)$ when
$\theta/\pi\in[0,0.182]$, and decreases with decreasing values of
both $D(B|A)$ and $D(C|A)$ when $\theta/\pi\in[0.818,1]$. As
illustrated in the inset of Fig. \ref{fig:2} with
$\theta/\pi\in[0,0.5]$, this counterintuitive phenomenon is caused
by the more quickly increasing rate of $D(C|A)$ (the dashed blue
line) compared with that of $D(B|A)$ (the solid red line). Out of
the above $\theta/\pi$ regions, either $D(B|A)$ increases more
rapidly than that of $D(C|A)$, or $D(B|A)$ increases while $D(C|A)$
decreases, and therefore the UB decreases with increasing $D(B|A)$.
So the behavior of UB depends on the competition of quantum
discords.

\subsection{Observation based on one-way unlocalizable quantum
discord}

Recently, Xi {\it et al.} proposed the concept of one-way
unlocalizable quantum discord \cite{Xizj}, which is in some sense
dual to quantum discord \cite{Ollivier}. Here, we present some
analysis of the quantum-memory-assisted entropic uncertainty
relation based on this measure of correlations. By using the same
semiological rules as Ref. \cite{Xizj}, we denote $E_a(\rho_{XY})$
as the entanglement of assistance for $\rho_{XY}$ \cite{Cohen},
while $E_u^\leftarrow(\rho_{XY})$ as the one-way unlocalizable
entanglement \cite{BGK}, and $\delta_u^\leftarrow(\rho_{XY})$ as the
one-way unlocalizable quantum discord \cite{Xizj}, both for
$\rho_{XY}$ with measurements on $Y$, and $X,Y\in\{A,B,C\}$. Then we
have the following result:

\emph{Proposition 2.} When the UB on $A$ is reduced with the aid of
a quantum memory $B$, then both
$E_u^\leftarrow(\rho_{BA})>E_u^\leftarrow(\rho_{CA})$ and
$\delta_u^\leftarrow(\rho_{BA})>\delta_u^\leftarrow(\rho_{CA})$ are
always satisfied.

\emph{Proof.} By using the Buscemi-Gour-Kim equality \cite{BGK}, we
have
\begin{eqnarray}\label{eq14}
 &&E_a(\rho_{BC})+E_u^\leftarrow(\rho_{BA})=S(\rho_B),\nonumber\\
 &&E_a(\rho_{CB})+E_u^\leftarrow(\rho_{CA})=S(\rho_C).
\end{eqnarray}
Substraction of the second equality of Eq. \eqref{eq14} from that of
the first one gives rise to
$E_u^\leftarrow(\rho_{BA})-E_u^\leftarrow(\rho_{CA})=S(\rho_B)-S(\rho_C)=-S(A|B)>0$,
and hence $E_u^\leftarrow(\rho_{BA})>E_u^\leftarrow(\rho_{CA})$.
Furthermore, combination of the definition of the one-way
unlocalizable quantum discord \cite{Xizj} with the Buscemi-Gour-Kim
equality \cite{BGK} implies
\begin{eqnarray}\label{eq15}
 &&\delta_u^\leftarrow(\rho_{BA})+S(B|A)=E_a(\rho_{BC}),\nonumber\\
 &&\delta_u^\leftarrow(\rho_{CA})+S(C|A)=E_a(\rho_{CB}).
\end{eqnarray}
Then we have
$\delta_u^\leftarrow(\rho_{BA})-\delta_u^\leftarrow(\rho_{CA})=S(C|A)-S(B|A)=-S(A|B)>0$,
and therefore
$\delta_u^\leftarrow(\rho_{BA})>\delta_u^\leftarrow(\rho_{CA})$.
\hfill{$\blacksquare $}

This proposition implies that when the UB on $A$ is reduced using
the information stored in a quantum memory $B$, then both the
one-way unlocalizable entanglement and the one-way unlocalizable
quantum discord between $A$ and $B$ are always larger than those
between $A$ and the purifying system $C$. This reinforces the
interpretation of the potential maximal correlations between $A$ and
$B$ as the essential element responsible for the reduction of the
measurement uncertainty in Eq. \eqref{eq1}.

\section{Negative conditional entropy}
As the negativity of the
conditional entropy plays such an important role in improving the
prediction precision of the uncertainty game, we now present some
possible structures of $\rho_{AB}$ ensuring $S(A|B)<0$. By noting
the Araki-Lieb inequality \cite{Nielsen}
\begin{eqnarray}\label{eq16}
 S(\rho_{AB})\geqslant |S(\rho_A)-S(\rho_B)|,
\end{eqnarray}
we see that if $S(\rho_B)-S(\rho_A)=S(\rho_{AB})$, then
$S(A|B)=-S(\rho_A)\leqslant 0$ due to the non-negativity of the von
Neumann entropy. $S(A|B)$ is negative if $S(\rho_{A})\neq 0$, i.e.,
$\rho_A\neq |\mu\rangle\langle\mu|$, with $|\mu\rangle$ being the
orthonormal basis of $\mathcal {H}_A$. Recently, a necessary and
sufficient equality condition for the inequality \eqref{eq16} was
derived in \cite{Wujd}. It states that
$S(\rho_B)-S(\rho_A)=S(\rho_{AB})$ if and only if the complex
Hilbert space $\mathcal {H}_B$ can be factorized as $\mathcal
{H}_B=\mathcal {H}_{B^L}\otimes \mathcal {H}_{B^R}$ such that
\begin{eqnarray}\label{eq17}
 \rho_{AB}=|\psi\rangle_{AB^L}\langle\psi|\otimes\rho_{B^R},
\end{eqnarray}
with $|\psi\rangle_{AB^L}\in\mathcal {H}_{A}\otimes\mathcal
{H}_{B^L}$.

In fact, for state $\rho_{AB}$ of Eq. \eqref{eq17}, we have
\begin{eqnarray}\label{eq18}
 S(\rho_{AB})&=&S(|\psi\rangle_{AB^L}\langle\psi|)+S(\rho_{B^R})\nonumber\\
             &=&S(\rho_{B^R})=S(\rho_B)-S(\rho_{B^L})\nonumber\\
             &=&S(\rho_B)-S(\rho_{A}),
\end{eqnarray}
by using the additivity of the von Neumann entropy \cite{Nielsen},
and therefore $D(B|A)=D(B_L|A)=S(\rho_A)$. Combination of this with
Eq. \eqref{eq9} gives $J(C|A)=0$. Since quantum correlation cannot
exist without classical correlation \cite{Ollivier}, this further
implies $D(C|A)=0$ and $J(B|A)=S(\rho_A)$, which confirms the
arguments presented in Proposition 1, namely, $D(B|A)>D(C|A)$ and
$J(B|A)>J(C|A)$ if $S(A|B)<0$.

As an explicit example, consider a qubit-qudit system with
$\rho_{AB}=(|00\rangle+|12\rangle)(\langle 00|+\langle
12|)/4+(|01\rangle+|13\rangle)(\langle 01|+\langle 13|)/4$ in the
standard basis $\{|\mu\nu\rangle\}_{\mu\nu=00}^{13}$. As shown in
Ref. \cite{Xi}, this state can be factorized as in Eq. \eqref{eq17}
with $|\psi\rangle_{AB^L}=(|00\rangle+|11\rangle)/\sqrt{2}$ and
$\rho_{B^R}=\mathbb{I}_{B^R}/2$, and as a result gives the negative
conditional entropy $S(A|B)=-S(\rho_A)=-1$.

Finally, note that Eq. \eqref{eq17} is only a
sufficient condition for the negativity of $S(A|B)$, and there are
bipartite states $\rho_{AB}$ ensuring $S(A|B)<0$, but cannot be
factorized into the form of Eq. \eqref{eq17}. An obvious example of
such states is the two-qubit Werner state
$\rho_{AB}=r|\Psi\rangle\langle\Psi|+(1-r)\mathbb{I}_4/4$
\cite{Werner}, with $|\Psi\rangle=(|00\rangle+|11\rangle)/\sqrt{2}$
and $r\gtrsim 0.7476$.

\section{Summary and discussion}
To summarize, we have established
some new physical implications of the quantum-memory-assisted
entropic uncertainty relation from the perspective of correlative
capacities, which are captured by the concepts of quantum discord,
EoF, and the one-way unlocalizable quantum discord. The changing of
the uncertainty bound is a result of competitions of various
correlations between different players. We showed that whenever the
prediction precision is improved compared with that with only
classical memory, the observed particle $A$ reaches its potential
maximal correlative capacities with the quantum memory $B$ in the
sense that their correlations (both quantum and classical) are
always larger than those between $A$ and the purifying system $C$.
We hope these results may shed some new light on exploring the
physical implications of the entropic uncertainty principle,
especially from the perspective of quantum correlations.

As a concluding remark, we point out that the resulting certainty on
the prediction of the measurement outcomes of two incompatible
observables with the aid of a quantum memory may imply another kind
of uncertainty. This is convinced by a variant of the ``uncertainty
game" with more than two players, e.g., the three-player case
illustrated in Fig. \ref{fig:1}, which shows that the more precisely
the measurement outcomes of Alice are predicted by Bob, the less
precisely that will be predicted by Charlie, and vice versa.

\section*{ACKNOWLEDGMENTS}
This work was supported by NSFC (11205121, 10974247, 11175248), the
``973'' program (2010CB922904), NSF of Shaanxi Province
(2010JM1011), and the Scientific Research Program of Education
Department of Shaanxi Provincial Government (12JK0986).

\newcommand{\PRL}{Phys. Rev. Lett. }
\newcommand{\PRA}{Phys. Rev. A }
\newcommand{\APL}{Appl. Phys. Lett. }
\newcommand{\JPA}{J. Phys. A }
\newcommand{\JPB}{J. Phys. B }
\newcommand{\PLA}{Phys. Lett. A }
\newcommand{\NP}{Nature Phys. }
%

% BibTeX users please use
% \bibliographystyle{}
% \bibliography{}

\end{document}